\documentclass{article}
\usepackage{amsmath}
\usepackage{graphics}
\usepackage{graphicx}
\usepackage{natbib}
\usepackage{algorithm}
\usepackage{caption}
\usepackage{subcaption}
\usepackage{microtype}
\usepackage[margin = 1.5in]{geometry}
\DeclareGraphicsExtensions{.pdf}
\graphicspath{{fig/}}

\title{Likelihood free inference for Markov processes: a comparison}
\author{Jamie Owen\footnote{email: j.r.owen@ncl.ac.uk} \and Darren J. Wilkinson \and Colin S. Gillespie}
\date{\vspace{-5ex}}
\begin{document}

\maketitle

\begin{center}
{School of Mathematics \& Statistics, Newcastle University, \\ Newcastle upon Tyne, NE1 7RU, UK}
\end{center}
\begin{abstract}
  \label{sec:abstract}
  Approaches to Bayesian inference for problems with intractable likelihoods
  have become increasingly important in recent years. Approximate Bayesian
  computation (ABC) and ``likelihood free'' Markov chain Monte Carlo techniques
  are popular methods for tackling inference in these scenarios but such techniques are
  computationally expensive. In this paper we compare the two approaches to
  inference, with a particular focus on parameter inference for stochastic
  kinetic models, widely used in systems biology. Discrete time transition kernels for models of this type are intractable
  for all but the most trivial systems yet forward simulation is usually straightforward. We discuss the relative merits and drawbacks of each approach
  whilst considering the computational cost implications and efficiency of these
  techniques. In order to explore the properties of each approach we examine a
  range of observation regimes using two example models. We use a
  Lotka--Volterra predator prey model to explore the impact of full or partial
  species observations using various time course observations under the
  assumption of known and unknown measurement error. Further investigation into
  the impact of observation error is then made using a Schl{\"o}gl system, a
  test case which exhibits bi-modal state stability in some regions of parameter
  space.
\end{abstract}


\section{Introduction}
\label{sec:intro} 

Computational systems biology is concerned with the development of dynamic simulation models for complex biological processes \citep{kitano2002computational}. Such models are useful for contributing to the quantitative understanding of the underlying process through in-silico experimentation that would be otherwise difficult, time consuming or expensive to undertake in a laboratory. Stochastic kinetic models describe the probabilistic evolution of a dynamical system made up of a network of reactions. Models of this type are increasingly used to describe the evolution of biological systems \citep{golightly2005bayesian,proctor2007modelling,boys2008bayesian,wilkinson2009stochastic}. Motivated by the need to incorporate intrinsic stochasticity in the underlying mechanics of the systems these models are naturally represented by a Markov jump process.
Such systems are governed by a reaction network, each of which changes the state by a discrete amount and are hence naturally represented by a continuous time Markov process on a discrete state space.  State transition densities for models of this type are analytically intractable but forward simulation is available through use of, for example, the Direct method described by \cite{gillespie1977exact}. Models typically have a number of rate parameters which are important and of interest, but inference for these is an extremely challenging problem.
 
Parameter inference for Markov process models is often a computationally demanding problem due in part to the intractable likelihood function. Exact inference is possible through particle Markov chain Monte Carlo (pMCMC) \citep{andrieu2009pseudo,andrieu2010particle}, computationally intensive methods that make use of sequential Monte Carlo sampling techniques, embedded in a MCMC scheme.
PMCMC in this context requires running a sequential Monte Carlo filter, such as a bootstrap particle filter at each MCMC iteration, to provide an estimate to the likelihood.
The bootstrap filter is dependent on multiple forward simulations from the model for reliable estimation, leading to an expensive algorithm.
 
Approximate Bayesian computation (ABC) techniques have also shown to be a useful development when tackling problems with intractable likelihood functions. They allow inference in this scenario via an approximation to the posterior distribution. As in pMCMC, the ABC framework depends on a large number of model realisations given proposed parameter vectors, retaining samples which yield simulated data that is deemed to be sufficiently close to the observed data set, see \cite{tavare1997inferring,pritchard1999population,beaumont2002approximate}. Using a simple rejection sampling approach often leads to poor acceptance rates for tolerance thresholds that give an accurate approximation to the posterior, \citep{pritchard1999population}, meaning that a potentially very high number of data simulations must be made in order to obtain a good sample. Advancements within this framework have led to MCMC schemes and sequential Monte Carlo schemes, which typically return better acceptance rates than the simple rejection sampler, \citep{marjoram2003markov,del2006sequential,sisson2007sequential,toni2009approximate}.

Both particle MCMC, as in \cite{golightly2011bayesian} and \cite{wilkinson2011stochastic}, and ABC methods, \citep{drovandi2011estimation,fearnhead2012constructing}, have been successfully applied in the context of stochastic kinetics, but it is unclear as to which approach is favorable. If exact posterior inference is desired we are limited to particle MCMC. However, if exact inference is not the primary concern and there are computational constraints, perhaps available CPU time, it is not obvious which approach should be employed. Is it the case that increased computational efficiency is an adequate trade--off for the reduction in accuracy? In this article we explore both approaches under a range of situations in an attempt to draw some preliminary conclusions on which inference scheme may perform most efficiently in the context of parameter inference for Markov processes. Efficiency will be considered under the restriction that we apply the notion of a computational budget on the allowed number of model realisations in order to make like for like comparisons. This is under the assumption that given infinite time, as well as other conditions to guarantee convergence, each of these approaches would yield the same target. The budget is set on the number of model realisations since it is often the case that the forward simulation constitutes the bulk of computational cost of algorithms of this type.

\section{Stochastic kinetic models}
\label{sec:skm}

Consider a network of reactions which involves a set of $u$ species ${\cal X}_{1},
\ldots, {\cal X}_{u}$ and $v$ reactions $\mathbf{R}_{1}, \ldots, \mathbf{R}_{v}$
where each reaction $\mathbf{R}_{i}$ is given by
\begin{equation}  \label{eq:reaction}
  \mathbf{R}_{i}: \quad p_{i,1}{\cal X}_{1} + \ldots + p_{i,u}{\cal X}_{u} \rightarrow q_{i,1}{\cal X}_{1} + \ldots + q_{i,u}{\cal X}_{u}.
\end{equation}
$p_{i,j}$ denotes the number of molecules of species ${\cal X}_{j}$ that will be consumed in reaction $\mathbf{R}_{i}$. Similarly $q_{i,j}$ are the number of molecules of ${\cal X}_{j}$ produced in the reaction. Letting $P$ be the $v \times u$ matrix of $p_{i,j}$'s and $Q$ the corresponding matrix of coefficients of products the reaction network can be summarised as
\begin{equation}
  P{\cal X} \longrightarrow Q{\cal X}.
\end{equation}
The stoichiometry matrix, defined
\begin{equation}
  S = (Q - P)^{\prime},
\end{equation}
is a useful way to encode the information of the network as its columns represent the change of state caused by the different reaction events. Define $X_{t} = (X_{1,t}, \ldots, X_{u,t})$ as the vector denoting the number of species ${\cal X}$ present at time $t$.

Each reaction $\mathbf{R}_{i}$ is assumed to have an associated constant $\theta_{i}$ and hazard function $h_{i}(X_{t},\theta_{i})$ which gives the propensity for a reaction event of type $i$ at time $t$ to occur. We can consider the hazard function as arising due to interactions between species in a well mixed population.
If $\theta = (\theta_{1}, \ldots. \theta_{v})$ and $h(X_{t},\theta) = (h_{1}(X_{t},\theta_{1}), \ldots, h_{v}(X_{t},\theta_{v}))$ then full specification of the Markov process is complete given values for $\theta$ and $X_{0}$.

Exact trajectories of the evolution of species counts of such a system can be obtained via the Direct method, \cite{gillespie1977exact}. Algorithm~\ref{alg:gillespie} describes the procedure for forward simulation of a stochastic kinetic model given its stoichiometry matrix, $S$, reaction rates $\theta$, associated hazard function $h(X_{t}, \theta)$ and initial state $X_{0}$. Reactions simulated to occur in this way incorporate the stochastic nature and discrete state space of the system as reactions are chosen probabilistically and modify the state by discrete amounts. Whilst the Direct method allows exact simulation of the time and type of each reaction event that occurs, observed data $\mathcal{D} = (d_{0},d_{1}, \ldots, d_{T})$ are typically noisy, possibly partial, observations at discrete time intervals,

\begin{equation}
  d_{t} \sim \pi(\cdot|X_{t},\sigma),
\end{equation}
where $\sigma$ are parameters associated with the measurement error that may also need to be inferred.

Less computationally expensive simulation algorithms such as the chemical Langevin equation (CLE) relax the restriction imposed by the discrete state space, \cite{gillespie2000chemical}. The stochasticity of the underlying mechanics of the system is retained but realisations of the evolution of species levels are approximate. However \cite{gillespie2014diagnostics} show that such approximate simulators are not necessarily appropriate in all cases and that ensuring that they yield good approximations over the parameter space can be a problem. We therefore restrict ourselves to using the Direct method for the purposes of this article. For a comprehensive introduction into stochastic kinetic modelling see \cite{wilkinson2011stochastic}.

\begin{algorithm}[!t]
  \caption{The Direct method \citep{gillespie1977exact}}
  \label{alg:gillespie}
  \begin{enumerate}
  \item Set $t=0$. Initialise the rate constants $\theta$ and initial states $X_{0}$.
  \item Calculate the hazard function $h(X_{t},\theta)$ and $h_{0}(X_{t},\theta)
    = \sum_{i}^{v} h_{i}(X_{t},\theta_{i})$.
  \item Set $t = t + \delta t$ where
    \[
    \delta t \sim \operatorname{Exp}(h_{0}(X_{t},\theta)).
    \]
  \item Simulate the reaction index $j \in (1,\ldots,v)$ with probabilities
    \[
    p_{j} = \frac{h_{j}(X_{t},\theta)}{h_{0}(X_{t},\theta)} \;.
    \]
  \item Set $X_{t+\delta t} = X_{t} + S[j]$ where $S[j]$ is the
    $j^\text{th}$ column of the stoichiometry matrix $S$.
  \item If $t < T$ return to 2.
  \end{enumerate}
  
\end{algorithm}

\section{Bayesian inference for models with intractable likelihoods}
\label{sec:bayesinf}

\subsection{Particle Markov chain Monte Carlo}
\label{sec:pmcmc}

\begin{algorithm}[t]
  \caption{Pseudo-marginal MCMC \citep{andrieu2009pseudo}} \label{alg:pmcmc}
  \begin{enumerate}
  \item Initialise with a random starting value $\theta \sim \pi(\theta)$.
  \item Propose a move to a new candidate $\theta^{\ast} \sim q(\theta^{\ast}|\theta)$.
  \item Based on $\theta^{\ast}$, compute an unbiased estimate of $\pi(\mathcal{D}|\theta^{\ast})$, $\hat{\pi}(\mathcal{D}|\theta^{\ast})$,
  \item Accept the move with probability
    \begin{equation}
      \min\left\{1, \frac{\hat{\pi}(\mathcal{D}|\theta^{\ast})\pi(\theta^{\ast})q(\theta|\theta^{\ast})}
        {\hat{\pi}(\mathcal{D}|\theta)\pi(\theta)q(\theta^{\ast}|\theta)} \right\},
    \end{equation}
    else remain at $\theta$.
  \item Return to 2.
  \end{enumerate}
\end{algorithm}

Suppose we are interested in $\pi(\theta|\mathcal{D})$ and that we wish to construct an MCMC algorithm whose invariant distribution is exactly this posterior. Using an appropriate proposal distribution we can construct a Metropolis Hastings algorithm to do this. However this is often impractical due to the likelihood term, $\pi(\mathcal{D}|\theta)$, being unavailable. \cite{andrieu2009pseudo} proposed a pseudo marginal approach to this issue. In order to overcome this problem of the intractable likelihood function, we replace this evaluation in the Metropolis Hastings acceptance ratio with a Monte Carlo estimate $\hat{\pi}(\mathcal{D}|\theta)$ leading to the algorithm as described in algorithm~\ref{alg:pmcmc}. Provided that $E[\hat{\pi}(\mathcal{D}|\theta)] =
\pi(\mathcal{D}|\theta)$ the resulting stationary distribution is exactly the desired target. Within the context of Markov processes it is natural to make use of sequential Monte Carlo techniques through use of a bootstrap particle filter, \cite{doucet2001sequential}, described for this context in algorithm~\ref{alg:bpf}. The bootstrap filter gives unbiased estimates and hence the resultant MCMC scheme is ``exact approximate''.

\begin{algorithm}[t]
  \caption{Bootstrap particle filter} \label{alg:bpf}
  At time $t$ we have
  a set of $N$ particles $\mathbf{X}_{t}^{\ast} = \{(x_{t}^{i}, \pi_{t}^{i})  i
  = 1, \ldots, N\}$. The filter assumes fixed parameters and so we drop the
  $\theta$ from our notation. $t \in (0,1,\ldots,T)$ for observed data with $T$
  discrete time-point observations.
  \begin{enumerate}
  \item Initialise at $\mathbf{X}_{0}^{\ast}$, a set of $N$ independent draws from our prior distribution on the state.
  \item At time $t$, suppose we have a sample $\mathbf{X}_{t}^{\ast} \sim \pi(\mathbf{X}_{t}|D_{1:t})$.
  \item Sample a set of indices for candidates for forward simulation, $I_{t}^{i}$ according to the weights $\pi_{t}$.
  \item Simulate forward from the model the chosen paths, $x_{t+1}^{i} \sim \pi(x_{t+1}^{i}|x_{t}^{I^{i}_{t}})$.
  \item Calculate weights, $w_{t+1}^{i} = \pi(d_{t+1}|x_{t+1}^{i})$, and normalise to set $\pi_{t+1}^{i} = \frac{w_{t+1}^{i}}{\sum_{j=1}^{N} w_{t+1}^{j}}$.
  \end{enumerate}
  Define $\hat{\pi}(d_{t}|D_{1:t-1}) = \frac{1}{N}\sum_{i = 1}^{N} w_{t}^{i}$, then $\hat{\pi}(D_{1:t}) = \prod_{t=1}^{T} \hat{p}(d_{t},D_{1:t-1})$.
\end{algorithm}

The scheme as described here is a special case of the particle marginal Metropolis Hastings (PMMH) algorithm described in \cite{andrieu2010particle} which can also be used to target the full joint posterior $\pi(\theta,\mathbf{X}|\mathcal{D})$. It was noted in \cite{andrieu2009pseudo} that the efficiency of this scheme was dependent on the variance of the estimated likelihoods. Increasing the number of particles $N$ yields estimates with a smaller variance at the expense of increased computation time. Optimal choices for $N$ were subject to interest in \cite{pitt2012some} and \cite{doucet2012efficient}. The former suggest that the variance of the log--likelihood estimates should be around 1 in order to be optimal, however the latter argue that the efficiency penalty is small for values between 0.25 and 2.25.

\subsection{Approximate Bayesian computation}
\label{sec:abc}
Approximate Bayesian computation (ABC) techniques have increased in use in recent years due to their applicability to inference for a posterior distribution, $\pi(\theta|\mathcal{D})$, for problems in which evaluation of the likelihood function, $\pi(\mathcal{D}|\theta)$, due to cost or analytical intractability, is unavailable. Such methods are typically computationally intensive due to their reliance on the ability to simulate realisations from the model.
Ideally, given a collection of parameter vectors, $\theta$, we would keep all vectors that gave rise to simulated data which is equivalent to our observed data set. In practice however, the probability that a candidate data set $\mathcal{D}^{\ast} = \mathcal{D}$ is almost 0. Hence an approximation to the target distribution is made through a collection of samples of parameters that lead to data simulation which is deemed to be sufficiently close to the observations. Simulated data, $\mathcal{D}^{\ast}$, is considered to be close if, for a given metric $\rho(\cdot)$, the distance between simulated and observed data is smaller than some threshold $\epsilon$. The simple rejection sampler is described in algorithm~\ref{alg:abcrs}

\begin{algorithm}
  \caption{ABC rejection sampler}
  \label{alg:abcrs}
  \begin{enumerate}
  \item Generate a candidate parameter vector $\theta^{\ast} \sim \pi(\theta)$.
  \item Simulate a candidate data set $\mathcal{D}^{\ast} \sim
    \pi(\mathcal{D}|\theta^{\ast})$.
  \item Calculate a measure of distance between the candidate data,
    $\mathcal{D}^{\ast}$, and the observed data $\mathcal{D}$,
    $\rho(\mathcal{D},\mathcal{D}^{\ast})$.
  \item Accept $\theta^{\ast}$ if $\rho(\cdot) < \epsilon$ for some predetermined, fixed, $\epsilon$.
  \item Go to 1.
  \end{enumerate}
\end{algorithm}
Instead of yielding the true posterior distribution, samples have the
approximate distribution $\nobreak{\pi(\theta\,\vert
  \,\rho(\mathcal{D},\mathcal{D}^{\ast}) < \epsilon)}$. Acceptance rates of ABC
algorithms are often improved by employing dimension reduction techniques on the
data. This approximation tends to the true target as $\epsilon \rightarrow 0$
when $\rho(\cdot)$ is a properly defined metric on sufficient statistics.
Further approximations are made when, as is often the case, sufficient
statistics are unavailable. In this situation one would choose a set of summary
statistics that it is hoped is informative about the data.
\cite{blum2013comparative} give a review of current techniques for choosing
summary statistics. Over the past decade numerous proposals to improve the
efficiency of ABC have been made. Favored schemes include the use of a
sequential Monte Carlo sampler, which seeks to address the issue of poor
acceptance rates through first allowing a high acceptance threshold and then
gradually reducing the tolerance to improve the approximation to the target
distribution. This algorithm is sequential in the sense that populations of
simulated points (particles) are generated at each stage. The sample $\pi(\theta
\,\vert\, \rho(\mathcal{D},\mathcal{D}^{\ast}) < \epsilon_{t-1})$ is then
exploited as the basis for a proposal distribution used to target $\pi(\theta \,
\vert\, \rho(\mathcal{D},\mathcal{D}^{\ast}) < \epsilon_{t})$.
\cite{douc2007convergence} showed that from the perspective of importance
sampling this reliance on previous populations to improve proposals is entirely
legitimate. Early approaches often used a geometric rate of decline for the
tolerances, however adaptive schemes based on the distribution of the distances
have been shown to work well \citep{drovandi2011estimation}. It is of note
however that consideration must be given to the criteria by which we choose the
new tolerance, as \cite{silk2013optimizing} showed that convergence is not
guaranteed in all cases. A sequential approach to inference within the ABC
framework based on importance sampling is described in
algorithm~\ref{alg:abcsmc}.

In practice there are a number of factors which contribute to the efficiency of the sequential scheme described here. A perturbation kernel $K_{t}(\cdot)$, typically some Gaussian distribution, with a small variance usually leads to good acceptance rates but slow exploration of the parameter space. Conversely larger moves will explore the space more quickly but at the cost of reduced acceptance probability. \cite{beaumont2009adaptive} consider use of an adaptive Gaussian proposal distribution, with  variance equivalent to twice the empirical variance of the samples, $\theta^{(t)}$. This was built on in an article by \cite{filippi2013optimality} who derived an optimal proposal variance, optimal in terms of jointly minimising the Kullback-Liebler divergence between proposal and target and maximising the acceptance rate, dependent on the current sample and the tolerance for the target. The sequence of tolerances and number of bridging distributions in the sequence also contribute to the overall effectiveness of the scheme. Intuitively, a slow decline in the threshold will lead to high acceptance rates for newly proposed parameter vectors, but posterior learning will be slow. 

\begin{algorithm}[t]
  \caption{Sequential ABC} \label{alg:abcsmc}
  \begin{enumerate}
  \item Initialise $\epsilon_{0} > \epsilon_{1} > \ldots\ > \epsilon_{T} > 0$ and set the population indicator, $t=0$.
  \item Set particle indicator, $i=1$.
  \item If t = 0, sample $\theta^{\ast \ast} \sim \pi(\theta)$ \\*
    Else sample $\theta^{\ast}$ from the previous population $\{\theta^{(i)}_{t-1}\}$ with weights $w_{t-1}$ and perturb to obtain $\theta^{\ast \ast} \sim K_{t}(\theta|\theta^{\ast})$ \\*
    If $\pi(\theta^{\ast \ast}) = 0$, return to 3. \\*
    Simulate a candidate dataset $x^{\ast} \sim f(x|\theta^{\ast \ast})$ \\*
    If $d(x_{0},x^{\ast}) \ge \epsilon_{t}$, return to 3.
  \item Set $\theta_{t}^{(i)} = \theta^{\ast \ast}$ and calculate weight for particle $\theta_{t}^{(i)}$, $w_{t}^{(i)}$
    \[
    w_{t}^{(i)} = \left\{ 
      \begin{array}{cl}
        1, & \text{if } t = 0 \\
        \frac{\pi(\theta_{t}^{(i)})}{\sum^{N}_{j=1} w_{t-1}^{(j)} K_{t}(\theta^{(j)}_{t-1},\theta^{(i)}_{t})}, & \text{if } t > 0
      \end{array} 
    \right\}.
    \]
    If $i < N$ set $i = i+1$, go to 3
  \item Normalise the weights, if $t<T$, set $t=t+1$ and go to 2.
  \end{enumerate}
\end{algorithm}

\section{Basis for comparison}
\label{sec:mfc}

In order to create a framework in which we can make meaningful comparisons between these approaches to inference it is important to consider what makes a fair test, as well as some measure of efficiency of each sampler. In addition to this we are interested in the discrepancy between the resultant posterior and the true posterior. One of the primary motivations for the comparison is to determine which method is most appropriate with particular consideration to the notion of a restricted computational budget.
In order to maintain as much consistency as possible over the various runs we use the Direct method, \cite{gillespie1977exact}, for all realisations from a given model. We shall compare a pseudo-marginal Metropolis Hastings implementation of pMCMC with ABC approximations that use a Euclidean metric function over the full set of data points,
\begin{equation}
  \label{eq:metric}
  \rho(\mathcal{D},\mathcal{D}^{\ast}) = \sum_{t = 0}^{T} \sum_{i = 0}^{U} (d^{\ast}_{i,t} - d_{i,t})^{2}.
\end{equation}
We make this choice to ensure that, in the limit, we are targeting the same posterior. Additionally, we have found that this metric performs competitively with other choices for the sample sizes considered in our simulations study.
We repeat runs of each algorithm for a range of observation schemes on a number of data sets for each model, each using the same computational budget, comparing the results from each. We also include for each a long run of a pMCMC scheme which will provide us with the `true' posterior of interest in each case.

\subsection{Computational budget}
\label{ssec:cb}

We define our computational budget by considering each model realisation via the Direct method as 1 computational unit. We ignore the contribution to computation time of all other aspects of each algorithm as typically it is the path simulation that takes up the majority of computation time. In addition this choice ensures that the comparisons made are unaffected by certain computational optimisations, and coding tricks, which may distort the results in favor of one algorithm over another. For example ABC typically parallelises trivially and often yields almost perfect scaling between number of processors used and the speed factor gained, whereas particle MCMC does not benefit from parallel hardware in the same way, but can still be parallelised.

\subsection{Initialisation}

Each of the approaches to inference described above exhibit aspects which need to be chosen in some way, each of which has a bearing on the efficiency of the sampler. To make comparison as fair as possible we want each algorithm to be in some sense optimised using standard published methods. We now explain, for clarity, the way in which we have chosen various tuning parameters for each of the algorithms above. The cost of obtaining such parameters is collected and is deducted from our computational budget.

\subsubsection{Particle MCMC}

It has been well documented that the efficiency of random walk Metropolis algorithms is highly dependent on the choice of proposal kernel. A distribution which yields small deviations from the current state will ensure that a large number of moves are accepted but samples will be highly correlated. Large moves around the space on the other hand will often be rejected leading to the chain spending large amounts of time stuck at the same value. Under various assumptions about the target it has been shown that the optimal scaling for a Gaussian proposal kernel is
\begin{equation}
  \Sigma_{q} = \frac{2.38^{2}}{\sqrt{d}}\Sigma
\end{equation}
where $\Sigma$ is the covariance of the posterior distribution and $d$ is the number of parameters being estimated, \citep{roberts1997weak, roberts2001optimal}.
The starting parameter vector, $\theta_{0}$, of the chain also has an effect on the efficiency of the sampler. A choice of $\theta_{0}$ which is far from a region of non--negligible posterior density will lead to a chain which takes a long time to move toward the target distribution, whereas a chain initialised close to stationarity will yield useful samples sooner. This burn--in period can sometimes consume a sizeable fraction of the computational budget.
Particle MCMC as described in section~\ref{sec:pmcmc} also relies on a sequential Monte Carlo algorithm for approximation of the likelihood, $\hat{\pi}(\mathcal{D}|\theta)$. The bootstrap filter requires multiple model realisations, via a set of ``particles'' in order to achieve this approximation. In addition the approximation has to be calculated at every iteration of the MCMC algorithm, hence clearly the number of particles in the filter will greatly affect the runtime of the resultant algorithm. A small number of particles will result in a shorter computation time for the likelihood approximation but leads to larger variability in the estimated likelihood. This increased variability leads to decreased efficiency of the inference scheme, as noted by \cite{andrieu2009pseudo}. A large number of particles, useful for consistent estimates of  $\hat{\pi}(\mathcal{D}|\theta)$ will lead to slower posterior sampling in the chain.
An optimal choice for the number of particles has been explored by \cite{pitt2012some} who suggest that the number of particles should be chosen such that the variance of the log--likelihood estimates is around 1. However \cite{doucet2012efficient} show that the efficiency of the scheme is good for variances between 0.25 and 2.25.

In practice, for the purpose of this comparison we choose an initial parameter vector, $\theta_{0}$, as a random sample from the posterior distribution. The number of particles used in the particle filter, $N$ is then chosen by repeated runs of a particle filter with increasing $N$ until $1.5 < \operatorname{Var}(\hat{l}(\theta_{0}|\mathcal{D})) < 1.8$. We then use the covariance matrix of the posterior, $\Sigma_{p}$ to inform our choice for $\Sigma_{q}$, the Gaussian random walk proposal variance,
\begin{equation}
  \Sigma_{q} = \frac{2.38^{2}}{\sqrt{d}}\Sigma_{p}.
\end{equation}

During our first experiments with the pMCMC algorithm for these models we approached initialisation and tuning of the algorithm under the assumption of no knowledge of the posterior distribution of interest. However, this proved to be problematic as finding a sensible choice of $\theta_{0}$, number of particles, $N$, and proposal variance, $\Sigma_{q}$, often used a large proportion of the allocated computational budget. Under the computational restrictions imposed by our budget choice this made pMCMC look completely uncompetitve. This problem itself is interesting as it highlights a potential drawback of using pMCMC in practice. We discuss this issue further in section~\ref{sssec:tuningproblem}. 

\subsubsection{ABC SMC}

Initialisation of a sequential ABC algorithm as described in section~\ref{sec:abc} is somewhat less involved. This is due to the fact that optimal Gaussian proposal kernels for advancement to subsequent targets can be calculated during execution. In addition the sequence of tolerances is chosen adaptively throughout the algorithm. It remains that there is need to specify an initial tolerance value, $\epsilon_{0}$. One could argue that tuning the choice of metric and summary statistics to be used is also of interest. Discussion of how one might do this is beyond the scope of this article however, since we are limiting ourselves to the choice in equation~\ref{eq:metric} so as to ensure that as $\epsilon \rightarrow 0$ the resultant posterior approximation tends toward the true posterior distribution of interest, $\pi(\theta|\rho(\mathcal{D},\mathcal{D}^{\ast}) < \epsilon) \rightarrow \pi(\theta|\mathcal{D})$. For an in depth discussion of methods in which to choose summary statistics see \cite{blum2013comparative}.
In order to choose a suitable $\epsilon_{0}$ for the scheme we simply calculate $\rho(\mathcal{D},\mathcal{D}^{\ast}|\theta)$ using a number of samples from $\pi(\theta)$. From this we take $\epsilon_{0}$ to be the value equivalent to the $1\%$-ile of the resultant distribution of distances.

\section{Case study}
\label{sec:cs}

\subsection{Lotka--Volterra predator prey model}
\label{ssec:lvppm}

\subsubsection{Model description}

The Lotka--Volterra predator prey system, \cite{lotka1925elements,volterra1926fluctuations} is an example of a stochastic kinetic model that provides an ample starting point for investigation of parameter inference in models of this type. Although it is
relatively simple, characterised by a set of 3 reactions on two species, it encompasses many of the difficulties associated with larger, more complex systems. Denoting the two species, prey, $X_{1}$ and predators, $X_{2}$, evolution of the system is governed by the following three reactions:
\begin{align}
  \centering
  \begin{array}{rccc}
    \mathbf{R}_{1}: & {\cal X}_{1} & \rightarrow & 2{\cal X}_{1} \\
    \mathbf{R}_{2}: & {\cal X}_{1} + {\cal X}_{2} & \rightarrow & 2{\cal X}_{2} \\
    \mathbf{R}_{3}: & {\cal X}_{2} & \rightarrow & \emptyset.
  \end{array}
  \label{eq:lvmodel}
\end{align}
These reactions can be thought of as prey birth, a predator prey interaction resulting in the death of a prey and a predator birth, and predator death respectively. Reaction events are dependent on the current state of the system as well as the reaction rate parameters. Hence the trajectory of the evolution of species counts presents a Markov process on a discrete state space. This reaction network is summarised by its stoichiometry matrix , $S$ and hazard function $h(\mathbf{X},\theta)$:
\begin{align}
  S &= \left(
    \begin{array}{rrr}
      1 & -1 & 0 \\
      0 & 1 & -1
    \end{array} \right), &
  h(\mathbf{X},\theta) &= (\theta_{1}X_{1}, \theta_{2}X_{1}X_{2}, \theta_{3}X_{2}).
\end{align}
Realisations from the model conditional on a vector of rate parameters, $\theta$, can be obtained exactly via algorithm~\ref{alg:gillespie}, or approximated via a number of fast simulation algorithms.

\subsubsection{Synthetic data}

\begin{figure}[!ht]
  \centering
  \includegraphics[width = \textwidth]{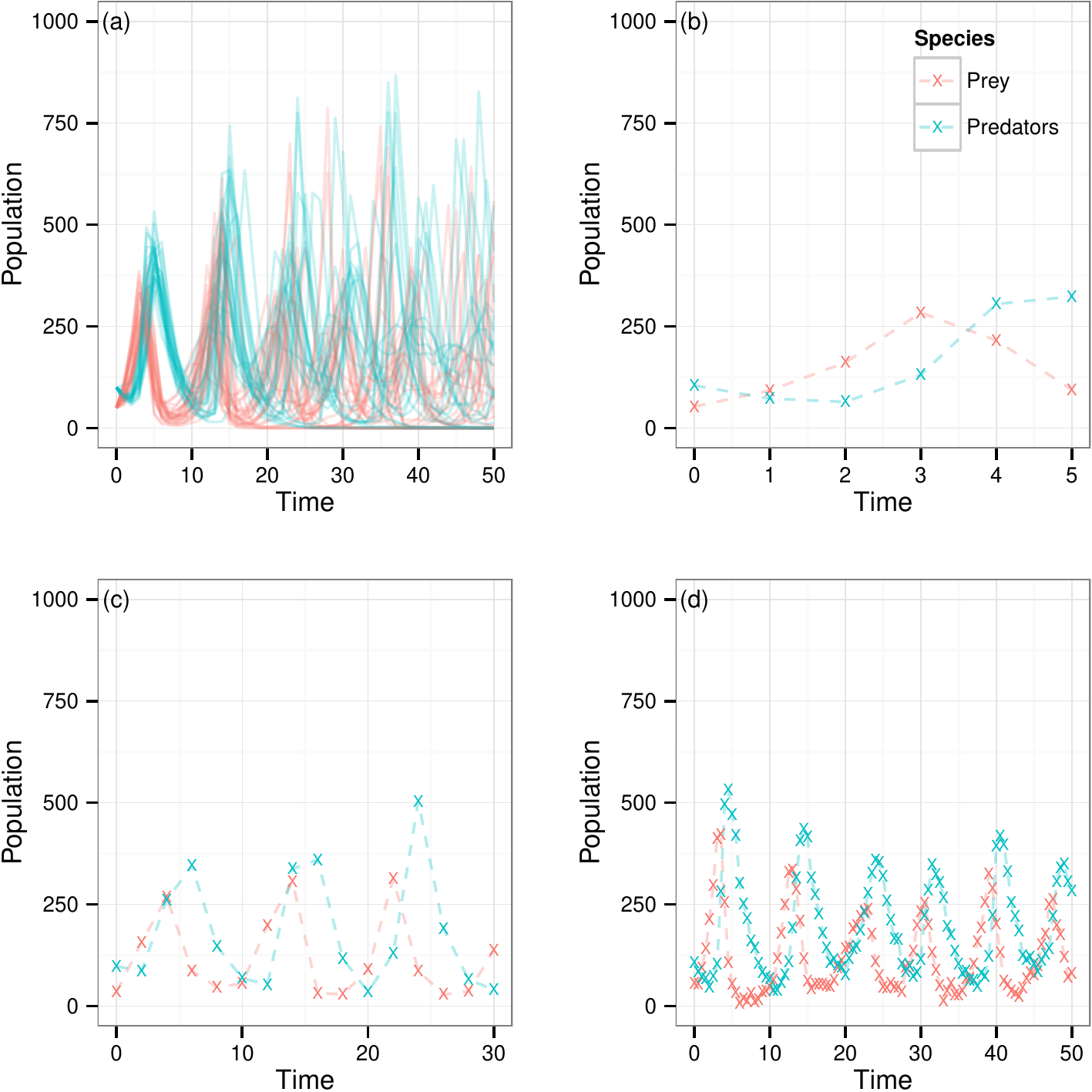}
  \caption{Synthetic data sets for the Lotka--Volterra predator prey model given $\log(\theta) = (0,-5.30,-0.51)$, $\log(\sigma) = 2.3$ and $X_{0} = (50,100)$. (a) are 25 realisations from the model using the Direct method, showing the oscillatory behavior and increasing volatility of the system. (b), dataset $\mathcal{D}^{1}$ is a short time series with observations at 6 time points at integer frequency. (c), $\mathcal{D}^{2}$: A time series observed at even time points with 16 time point measurements. (d), $\mathcal{D}^{3}$ we have a long time series of 101 time point measurements observed every 0.5 time units. We consider time series of differing lengths to determine whether the amount of data available has an influence on which inference method may be most appropriate. In each case we also consider a partial observation regime where predator observations are unavailable by discarding these measurements.}
  \label{fig:lvdata}
\end{figure}

For the purpose of making comparison we use a number of data sets over different observation regimes simulated using reaction rate vector $\theta = (1.0,0.005,0.6)$. In each case we corrupt the $X_{t}$ with a Gaussian error with mean 0 and variance $\sigma^{2}$, $\pi(d_{t}|X_{t}, \sigma) \sim \mathcal{N}(X_{t},\sigma^{2})$. $X_{0} = (50,100)$ is used throughout. Plots of each of the data sets considered are in figure~\ref{fig:lvdata}. Given this set of parameter values the model exhibits relatively stable oscillatory behavior for both species and provides an interesting starting point for our investigation. We shall use this model to explore posterior sampling efficiency given data sets of a range of sizes, under full and partial observation regimes, whilst also giving consideration to the effect of assuming known measurement or including this parameter in the set to be inferred. Data sets shown in figures~\ref{fig:lvdata}b--d are denoted $\mathcal{D}^{1}$, $\mathcal{D}^{2}$ and $\mathcal{D}^{3}$ respectively. We introduce extra subscript notation such that $\mathcal{D}^{1}_{p}$ implies the data set $\mathcal{D}^{1}$ where predator observations have been discarded and $\mathcal{D}^{1}_{u}$ symbolises treatment of $\mathcal{D}^{1}$ under the assumption of unknown measurement error. In addition $\mathcal{D}^{1}_{\ast}$ will be used as a reference to the collection of data sets $\mathcal{D}^{1}, \mathcal{D}^{1}_{u}, \mathcal{D}^{1}_{p}$ and $\mathcal{D}^{1}_{u,p}$.

\subsubsection{Inference set up}
We now create a scenario in which prior parameter information is poor. We place uniform prior information on $\log(\theta)$,
\begin{equation}
  \label{eq:lvprior}
  \log(\theta_{i}) \sim \mathcal{U}(-6,2), \quad i = 1,2,3,
\end{equation}
and we place a Poisson prior distribution on the initial state
\begin{equation}
  \label{eq:lvprior3}
  X_{1} \sim \operatorname{Pois}(50), \quad X_{2} \sim \operatorname{Pois}(100).
\end{equation}
Where $\sigma^{2}$ is not assumed to be known we use
\begin{equation}
  \label{eq:lvprior2}
  \log(\sigma) \sim \mathcal{U}(\log(0.5),\log(50)).
\end{equation}

For each repeat we allow a computational budget of $10^8$ model realisations from the Direct method, algorithm~\ref{alg:gillespie}. We choose this budget based on the fact that given the $\theta = (1.0,0.005,0.6)$ our simulator achieves $10^4$ simulations of length equivalent to $\mathcal{D}^{2}$ every 45-50 secs on our relatively fast Intel core i7-2600 clocked at 3.4 GHz. This yields an approximate total time spent simulating from the model of 14 hours plus some other comparably negligible computation costs for each individual inference run. 
Clearly improvement on simulation time can be made by parallelising the simulation of independent realisations from the direct method as well as other computational savings being made by clever optimisations in each algorithm. We have tried to disclude the effect of such algorithmic optimisation in our comparison as discussed in section~\ref{sec:mfc}. We include the information on approximate time here as a rough guide to practical implementation of inference for these types of models as well as the reasoning behind our particular budget choice.

\subsubsection{Discussion of results}
\label{sssec:lvresults}

\begin{figure}[!ht]
  \centering
  \includegraphics[width = \textwidth]{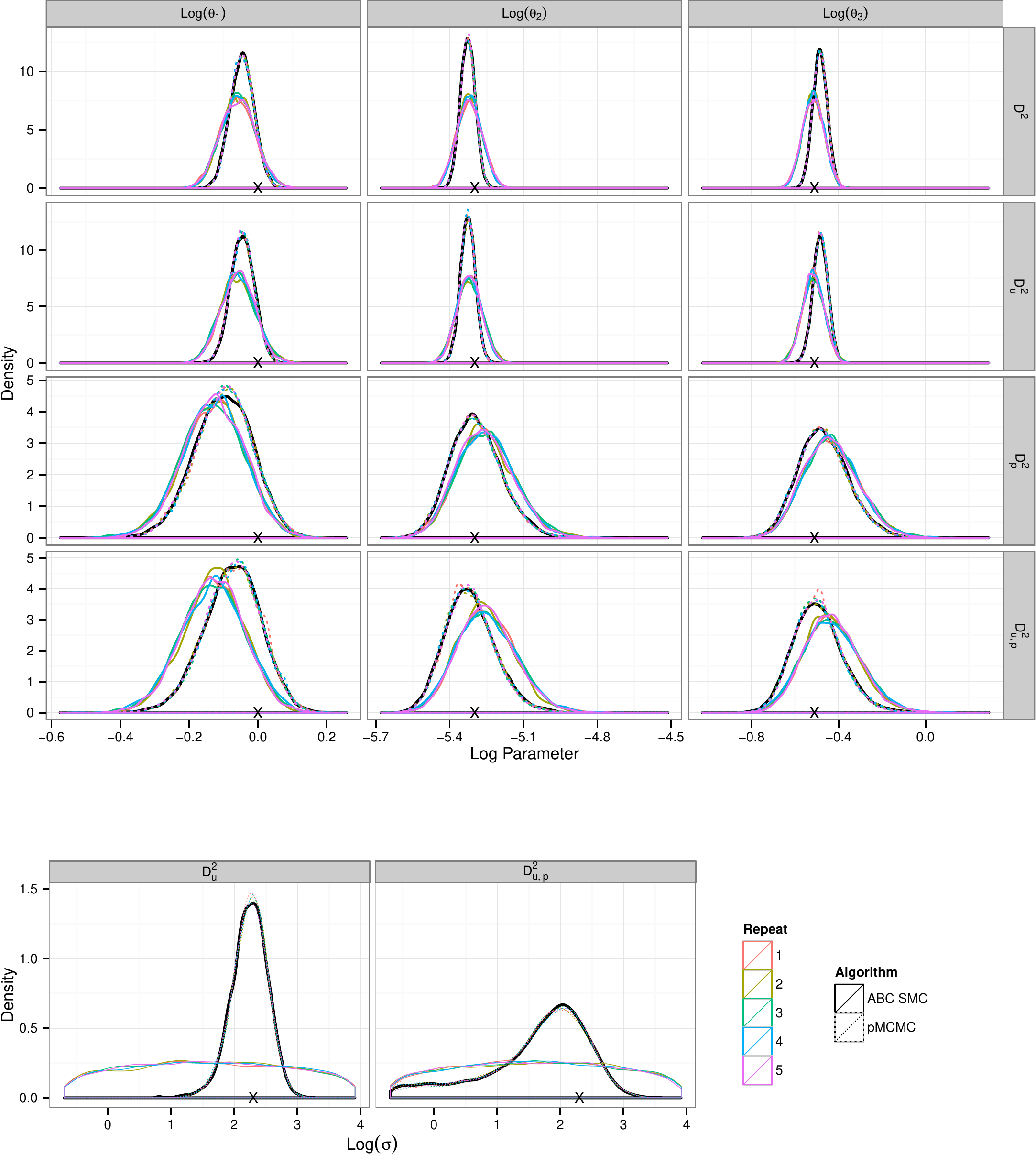}
  \caption{Posterior distributions given 5 repeats for each of the observation regimes using the $\mathcal{D}^{2}_{\ast}$ collection of data sets. True values are marked on the x-axis and a long pMCMC run with high number of particles to be used as a reference to the truth are in black.}
  \label{fig:lvinference}
\end{figure}

\begin{figure}[!ht]
  \centering
  \includegraphics[width = \textwidth]{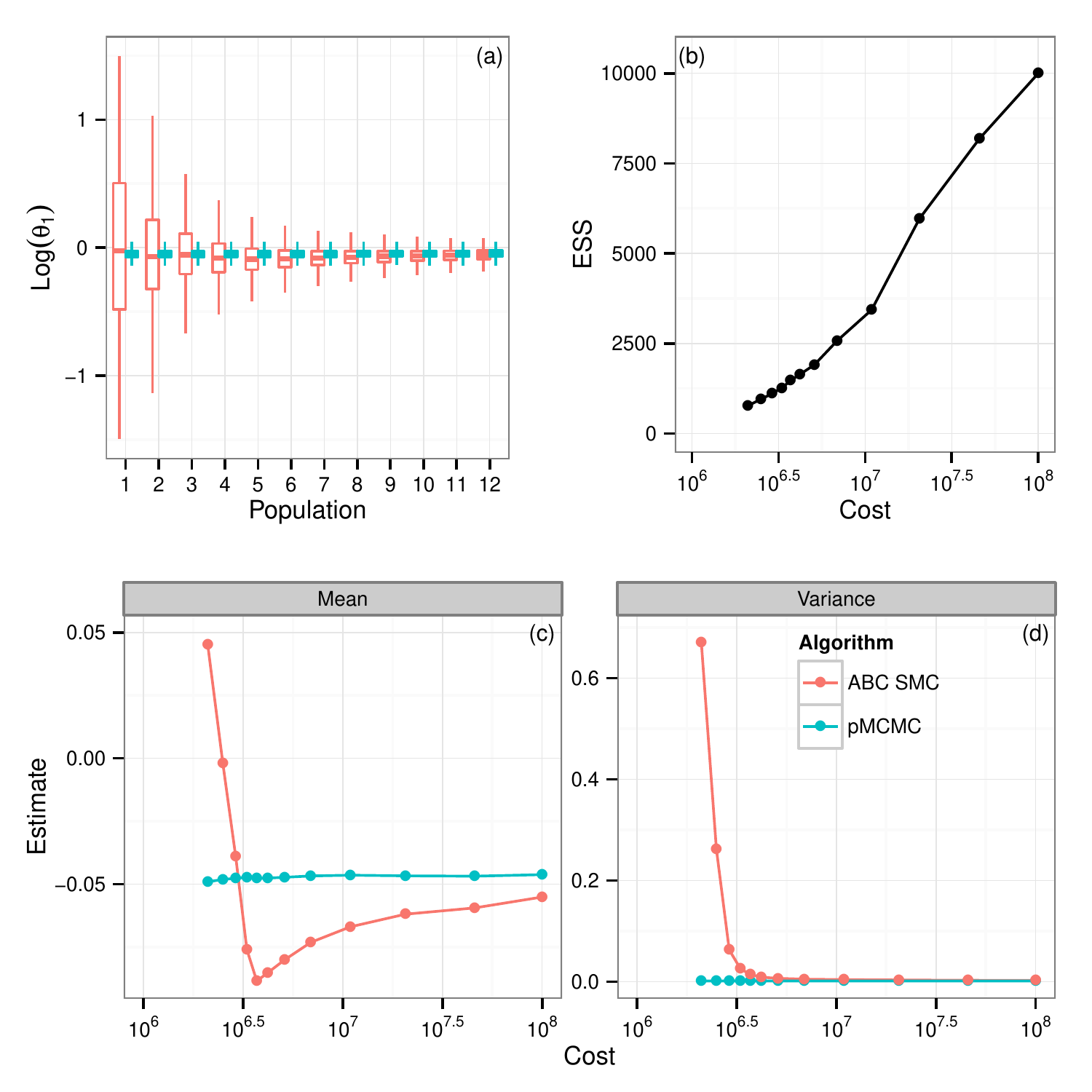}
  \caption{Box plots, (a), showing the posterior learning for each of the algorithms broken down by computational units. Each posterior sample through the sequence using ABC SMC is shown, with corresponding pMCMC inferences. This gives insight into how the two algorithms compare throughout the experiment. (b) shows the effective sample size of the pMCMC sample when broken into these computational groups, (c-d) are posterior estimates of the mean and variance respectively given the two algorithms. These results are for $\theta_{1}$ given one run of each of the algorithms. It is clear the ABC posterior distribution is tending toward the pMCMC posterior. The shape of the posterior distribution inferred using pMCMC does not change much throughout computation.}
  \label{fig:lvsummary}
\end{figure}

Results for data sets $\mathcal{D}^{1}_{\ast}$ and $\mathcal{D}^{3}_{\ast}$ are contained as supplementary material. We report results for data set $\mathcal{D}^{2}_{\ast}$ , figure~\ref{fig:lvdata}(c), shown in figure~\ref{fig:lvinference}.
Results in the supplementary material support those reported here, discrepancies between the two algorithms are reduced for $\mathcal{D}^{1}_{\ast}$ and exacerbated given the longer time series of $\mathcal{D}^{3}_{\ast}$. The plots in figure~\ref{fig:lvinference} show that for $\mathcal{D}^{2}$ there is a clear difference in the tails of the distribution. This feature is mirrored for $\mathcal{D}^{2}_{u}$. Additionally we see that, under the assumption of unknown measurement error ABC fails to identify the noise parameter, $\sigma$. Treatment of $\mathcal{D}^{2}_{p}$ yields a much smaller difference in the resultant posteriors between the two schemes however $\mathcal{D}^{2}_{u,p}$ reinforces the inability to infer $\sigma$ using the ABC SMC scheme. Each density plotted is the result of $10^4$ samples being collected. In the case of the ABC SMC this was ensured by retaining $10^4$ samples at each population, giving rise to a sequence of between 12 and 14 populations in the fully observed runs and 12 populations in each partially observed run, the decreasing choice of $\epsilon$ chosen as described in section~\ref{sec:abc}. For pMCMC we ran the sampler for the full budget and then thinned the resultant collection of vectors such that the final sample contained 10000. In the case of $\mathcal{D}^{2}$ the average number of particles required was 132 which led to a chain that ran for approximately $7.5 \times 10^{5}$ unthinned iterations which was then thinned by a factor of 75 to give the final sample.

From these results it would appear that given the use of the full budget pMCMC provides the better choice. Figure~\ref{fig:lvsummary}(a) shows the posterior learning experienced by the sequential ABC algorithm. The  matching box plots for pMCMC is the posterior distribution using the pMCMC algorithm where we use only information gained under the same budget use as with ABC. i.e The budget used to obtain the first population of samples under ABC was $2.1 \times 10^{6}$. The corresponding box-plot for population 1 under pMCMC is a snapshot of the chain having used up the same budget of $2.1 \times 10^{6}$. It is interesting to note that the shape of the posterior distribution under pMCMC changes little over the sequence of populations leading one to believe that even with little computational expense the posterior distribution inferred by pMCMC is good. However this plot does not tell the whole story. Figure~\ref{fig:lvsummary}(b) shows estimates of the effective sample size of the posterior distributions for pMCMC in this sequence. Effective sample size is small for the populations with the lower computational expense showing that to obtain a good posterior sample from pMCMC, despite the fact that the overall shape of the distribution doesn't change very much, we must still run for a long time, which is unsurprising. However, if we are only interested in obtaining posterior summaries, it would appear that the budget is less important. Figure~\ref{fig:lvsummary}(c-d) show posterior estimates of the mean and variance factored into the same computational expense groups as with figure~\ref{fig:lvsummary}(a) and figure~\ref{fig:lvsummary}(b). It is clear to see that for pMCMC the estimates are stable and remain reasonably constant. This is in stark contrast to estimates using the ABC approximations which appear less certain before tending toward those estimates gained using pMCMC.
The overall trend here then seems to suggest that in this instance pMCMC is the favorable choice. The posterior estimates of mean and variance are stable even given a relatively short run time, and the shape of the distribution is also maintained. To obtain a large uncorrelated sample the chain must be run for a long time, although it is noted that running the ABC sampler for the same length of time does not yield better results. Posterior summaries appear consistent irrespective of the additional runtime. ABC here however is not so good. It could be argued that the posterior distributions are close given the full budget in some circumstances, notably $\mathcal{D}^{2}_{p}$, however with shorter runtime the approximation is much greater and hence inference is poorer as a result.

\subsubsection{The tuning problem}
\label{sssec:tuningproblem}

The results presented here have ignored the issue of tuning the pMCMC algorithm. The posterior inference suggest that pMCMC is the better choice for learning about model parameters but we have tuned the pMCMC at the start, relying on knowledge of the posterior. This knowledge comes with its own expense and is in some sense self-defeating, something that is not so much of an issue for the ABC SMC. The true cost of the pMCMC inference then is somewhat higher than shown here. The ABC SMC proposal variance and tolerance sequence are chosen adaptively and so the initialisation and tuning cost is small. We made the choice to tune pMCMC using information from the posterior earlier due to the fact that in our initial experiments for this model we found that in a number of cases the computational budget used in initialising the pMCMC was large, often larger than our allocated budget. Choosing a sensible $\theta_{0}$, $N$ and $\Sigma_{q}$ with little prior knowledge is difficult.
Our initial attempts to find a $\theta_{0}$ involved sampling from the prior distribution, estimating the likelihoods and choosing the parameter vector which maximises the likelihood estimates. Due to the variability in the particle filter estimates away from the true parameter values this step typically involves a large number of particles and hence large expense. Conditional on this hopefully informative choice of $\theta_{0}$ we can then tune $N$, the number of particles in the bootstrap filter, such that each estimate has less expense, by running a number of filters starting with a small number of particles and steadily increasing $N$ until the $\operatorname{Var}(\hat{l})$ is suitably small. Again this has non-negligible expense. We want to try to find as small a value of $N$ as we can get away with for the main inference run and so this iterative procedure can be time consuming. On top of these two tuning steps we typically want to ensure that we choose a good $\Sigma_{q}$. This often involves a pilot MCMC run using a very small proposal variance and then using the variance of the resulting distribution to inform the choice of $\Sigma_{q}$.
We found that, in practice, employing these steps to tune pMCMC was itself very expensive. Using 1000 particles in a particle filter to estimate likelihoods for 2000 parameter vectors drawn from the prior for $\mathcal{D}^{2}$ and repeating each 10 times and maximising over the average to choose $\theta_{0}$ was not enough to guarantee that the resulting draw had posterior support. This alone uses 20\% of the allocated budget without then tuning the number of particles. The number of particles needed to satisfy the log--likelihood estimate variance criteria is often much larger in the tails of the posterior than at the true values. Add to this estimating $\Sigma_{q}$ and then the appropriate burn--in period and it is easy to see that this operation becomes very expensive.
 \cite{owen2014scalable} showed that initialising pMCMC with a random draw from an uninformative prior does not guarantee convergence to the stationary distribution. This was due to the large variability in log-likelihood estimates given by the particle filter in regions of negligible posterior support, rather than any flaw in the theory. It is suggested that a good thing to do in situations in which prior knowledge is poor is to run an ABC SMC algorithm as an aid to tuning and initialising a pMCMC algorithm. This approach is amenable to parallelisation and exploits relative strengths of the two approaches.

\subsection{Schl\"{o}gl system}
\label{ssec:schlogl}

\begin{figure}[!ht]
  \centering
  \begin{subfigure}{\textwidth}
    \includegraphics[width = \textwidth]{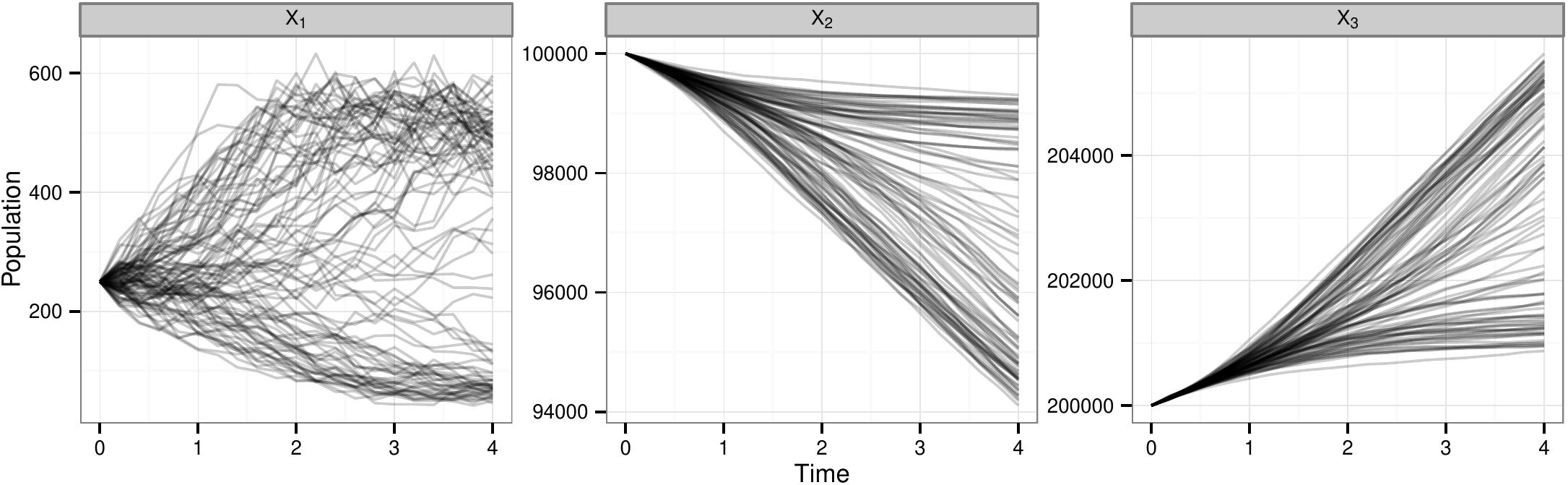}
    \caption{100 realisations from the Schl\"{o}gl system given highlight the bimodal stability shown for species $X_{1}$.}
    \label{fig:schloglsimulations}
  \end{subfigure}
  \\
  \begin{subfigure}{\textwidth}
    \includegraphics[width = \textwidth]{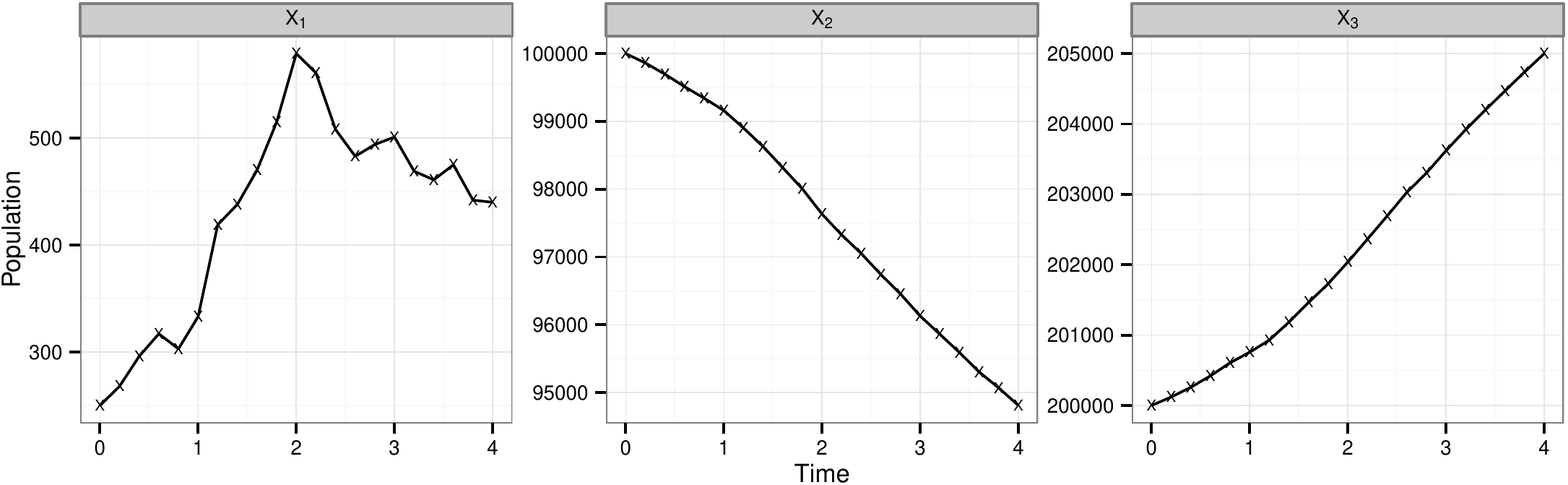}
    \caption{A pseudo data set to be used for inference where observations are made under Gaussian error, $\sigma^{2} = 1$}
    \label{fig:schlogldata}
  \end{subfigure}
  \caption{Simulating traces of the species present in the Schl\"{o}gl system using the Direct method with $\theta = (3\times10^{-7}, 10^{-4}, 0.000773, 3.276)$ and $X_{0} = (250,10^{5},2\times 10^5)$ recorded for 21 observations at regular intervals of 0.2 time units.}
\end{figure}

\subsubsection{Model description}
\label{sssec:smd}

The Schl\"{o}gl model is a well known test model which exhibits bimodal stability at certain parameter values. The system is characterised by a set of 4 reactions involving 3 species:

\begin{align}
  \centering
  \begin{array}{rccc}
    \mathbf{R}_{1}: & 2{\cal X}_{1} + {\cal X}_{2} & \rightarrow & 3{\cal X}_{1} \\
    \mathbf{R}_{2}: & 3{\cal X}_{1}  & \rightarrow & 2{\cal X}_{1} + {\cal X}_{2} \\
    \mathbf{R}_{3}: & {\cal X}_{3} & \rightarrow & {\cal X}_{1} \\
    \mathbf{R}_{4}: & {\cal X}_{1} & \rightarrow & {\cal X}_{3}. 
  \end{array}
  \label{eq:schloglmodel}
\end{align}

We can summarise this reaction network via its stoichiometry matrix $S$ and hazard function $h(\mathbf{X},\theta)$:
\begin{align}
  & \quad S = \left(
    \begin{array}{rrrr}
      1 & -1 & 1 & -1 \\
      -1 & 1 & 0 & 0 \\
      0 & 0 & -1 & 1
    \end{array} \right), \nonumber\\ \nonumber\\
  h(\mathbf{X},\theta) &= (\frac{\theta_{1}X_{1}(X_{1} - 1)X_{2}}{2}, \frac{\theta_{2}X_{1}(X_{1} - 1)(X_{1} - 2)}{6}, \theta_{3}X_{3}, \theta_{4}x_{1}).
\end{align}

This system provides an interesting case for which to investigate how influential the measurement error is on the efficiency of posterior sampling for each algorithm.

\subsubsection{Synthetic data}
\label{sssec:schlogldata}

Figure~\ref{fig:schloglsimulations} shows 100 realisations from the model given $\theta = (3\times 10^{-7}, 1\times 10^{-4}, 0.000773, 3.276)$ and $X_{0} = (250,10^5,2 \times 10^5)$ highlighting the bimodal characteristics given this set of parameter values. This poses an interesting challenge when it comes to parameter inference since it is not necessarily the case that all parameter vectors in a region of space closely surrounding this give the same behaviour. For our investigation of the inference methods being discussed we choose one of these data traces at random and corrupt with Gaussian error, $d_{t} \sim \mathcal{N}(X_{t},\sigma^{2})$, $\sigma^{2} = 1$. A second copy of the same underlying trace is then corrupted with the same error distribution but with $\sigma^{2} = 10$. We denote these data set $\mathcal{D}^{S}_{1}$ and $\mathcal{D}^{S}_{10}$ respectively. The chosen observed data set is shown in figure~\ref{fig:schlogldata}.

\subsubsection{Inference set up}
\label{sssec:schloglsetup}

In contrast to the simulation study for the Lotka--Volterra model we make use of a set of more informative prior distributions,
that is a Gaussian distribution on the log scale, centered at the true values with relatively small standard deviation, $0.5$. In addition we assume knowledge of the initial count $\mathbf{X}_{0}$. For inference in this case we now restrict our focus to assuming known measurement error, and availability of observations of all 3 species in the model, these factors having been explored well in the previous example. The focus of this example is to determine whether the size of the measurement error informs our choice as to which algorithm is better.

\subsubsection{Discussion of results}

\begin{figure}[!ht]
  \includegraphics[width=\textwidth]{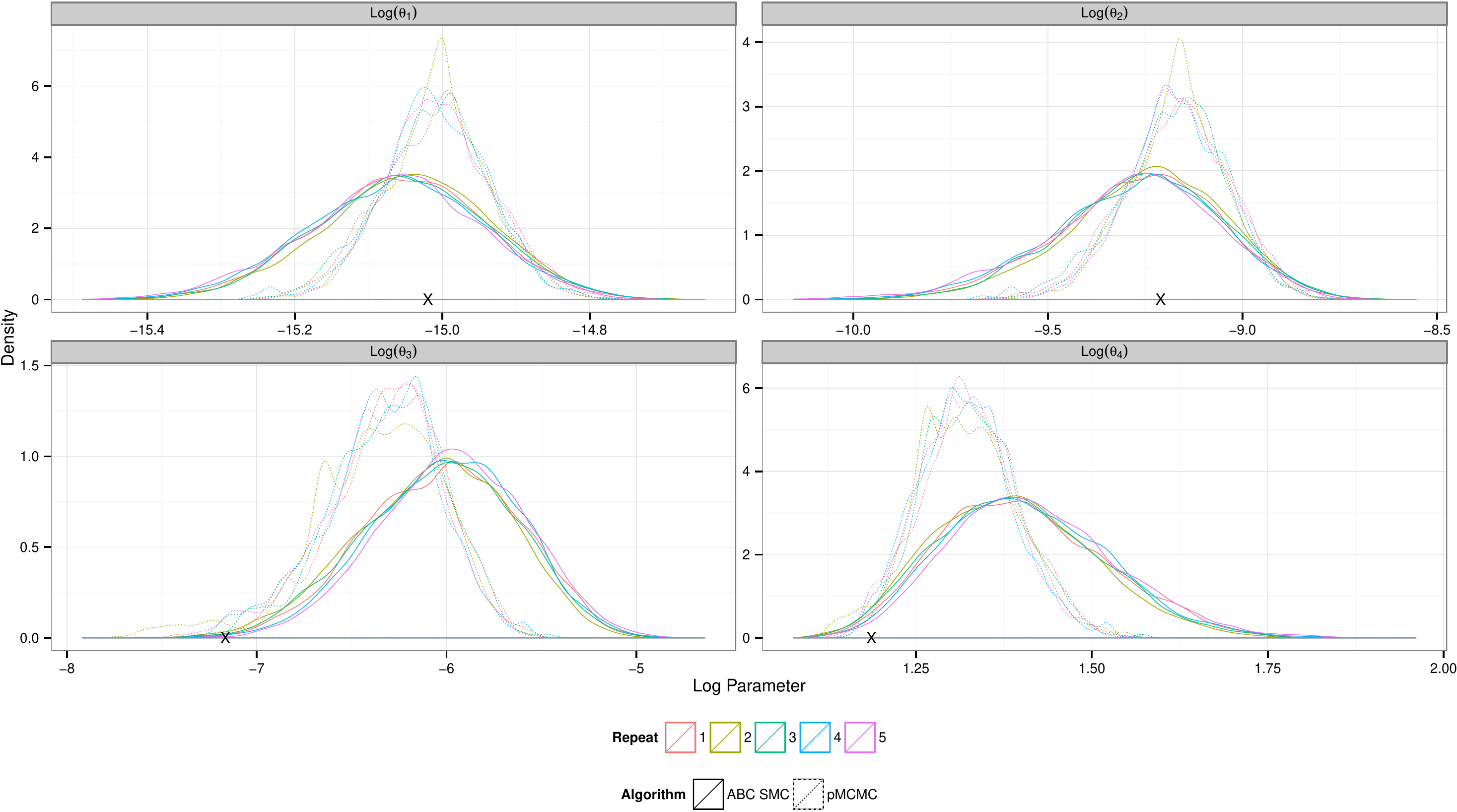}
  \caption{Posterior distributions for each of the 4 rate parameters over the the replications for the two algorithms given data set $\mathcal{D}^{S}_{10}$. ABC SMC over estimates the variance.}
  \label{fig:schlogllarge}
\end{figure}

\begin{figure}[!ht]
\includegraphics[width = \textwidth]{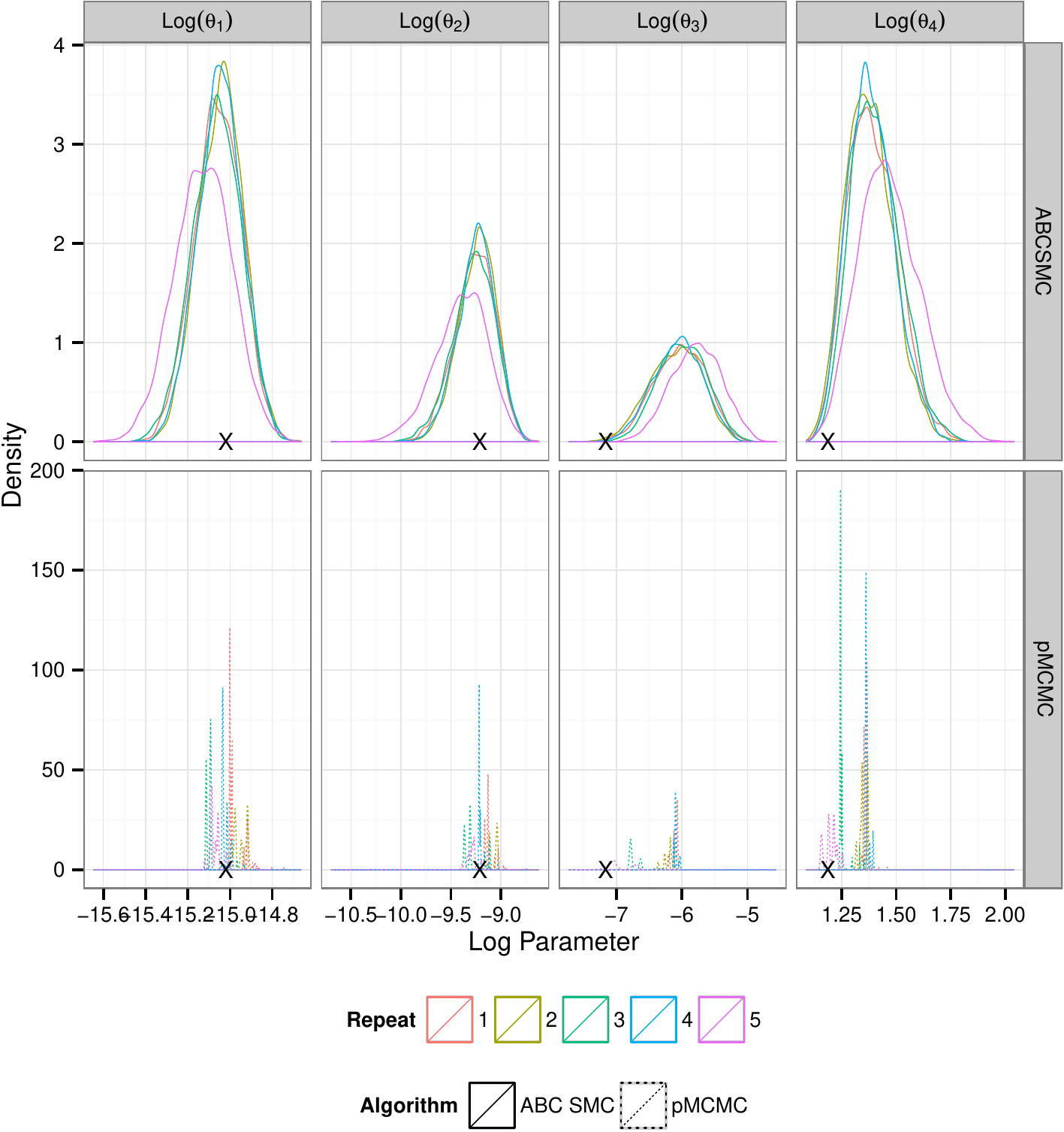}
\caption{Posterior distributions for each of the 4 rate parameters over the replications for the two algorithms given data set $\mathcal{D}^{S}_{1}$. PMCMC performs poorly when measurement error is small whereas inference using ABC appears to be good.}
\label{fig:schloglsmall}
\end{figure}

Consistent with the results in section~\ref{sssec:lvresults} ABC SMC yields posterior distribution with a larger mass in the tails. The plots in figure~\ref{fig:schlogllarge} show that under the larger measurement error, data set $\mathcal{D}^{S}_{10}$, comparative performance is similar to that seen in the previous example. However under small measurement error pMCMC struggles with this computational budget. The chain spends a large amount of time stuck at given parameter values. It is known that for likelihood free pMCMC to be efficient in this context the measurement error must be substantial. Development of pMCMC algorithms for informative observations are subject to ongoing research and typically involve bridging of the latent state conditional on the endpoints. \cite{golightly2014bayesian} propose an approach to the problem of inference for a Markov jump process with informative observations on a discrete state space by conditioning the hazard function on the end points of the data observations.

\section{Conclusions}
\label{sec:cr}

The results in this article suggest that in most cases, for parameter inference for stochastic kinetic models with intractable likelihoods, particle MCMC is a better choice than ABC SMC provided that it can be well initialised. This distinction is less clear when applied to small data sets as seen in appendix~\ref{sec:supplementary} where posterior inference for the rate parameters are well matched. A longer time series highlights the benefit of using a particle filter whose re-sampling step ensures forward simulations are guided by the observations. ABC SMC seems to be poor at inferring the measurement error present in the model but proves to be a reasonable approach under informative observations, favorable to the likelihood free pMCMC implementation explored here.
PMCMC, whilst being the better choice in most cases for inference is substantially more difficult to tune. Given that ABC methods parallelise easily and performs comparably when inferring rate parameters it poses a strong approach when measurement error is known or small, particularly for lower dimensional data and may yield good posterior distributions at a lower real time cost than pMCMC. The biggest trade-off here is finding an appropriate starting point for pMCMC, something that can be approached by using ABC SMC, as described in \cite{owen2014scalable}.

\bibliography{references}
\bibliographystyle{jfm}

\appendix
\section{Supplementary material}
\label{sec:supplementary}

\subsection{Analysing Lotka--Volterra $\mathcal{D}^{1}_{\ast}$}

\begin{figure}[!h]
  \centering
  \includegraphics[width = \textwidth]{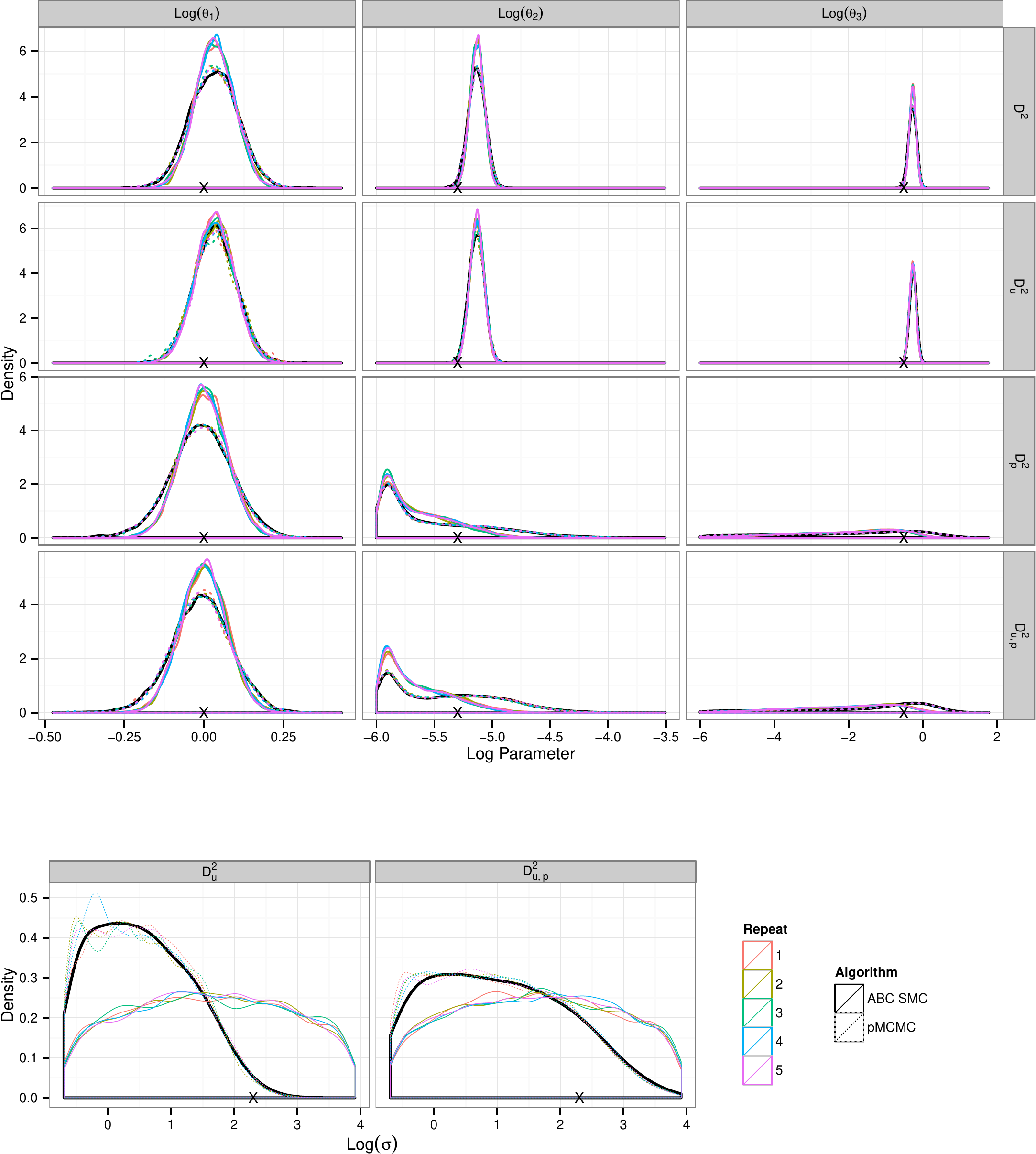}
  \caption{Posterior distributions over 5 repeats using each of the pMCMC and ABC SMC algorithms for the $\mathcal{D}^{1}_{\ast}$ collection of data sets. Densities are closely matched for the rate parameters in each case, however inference for $\sigma$ using ABC is poor. These results are consistent with those found in the text in section~\ref{sssec:lvresults}.}
  \label{fig:lvsmallposterior}
\end{figure}

\begin{figure}[!h]
  \centering
  \includegraphics[width = \textwidth]{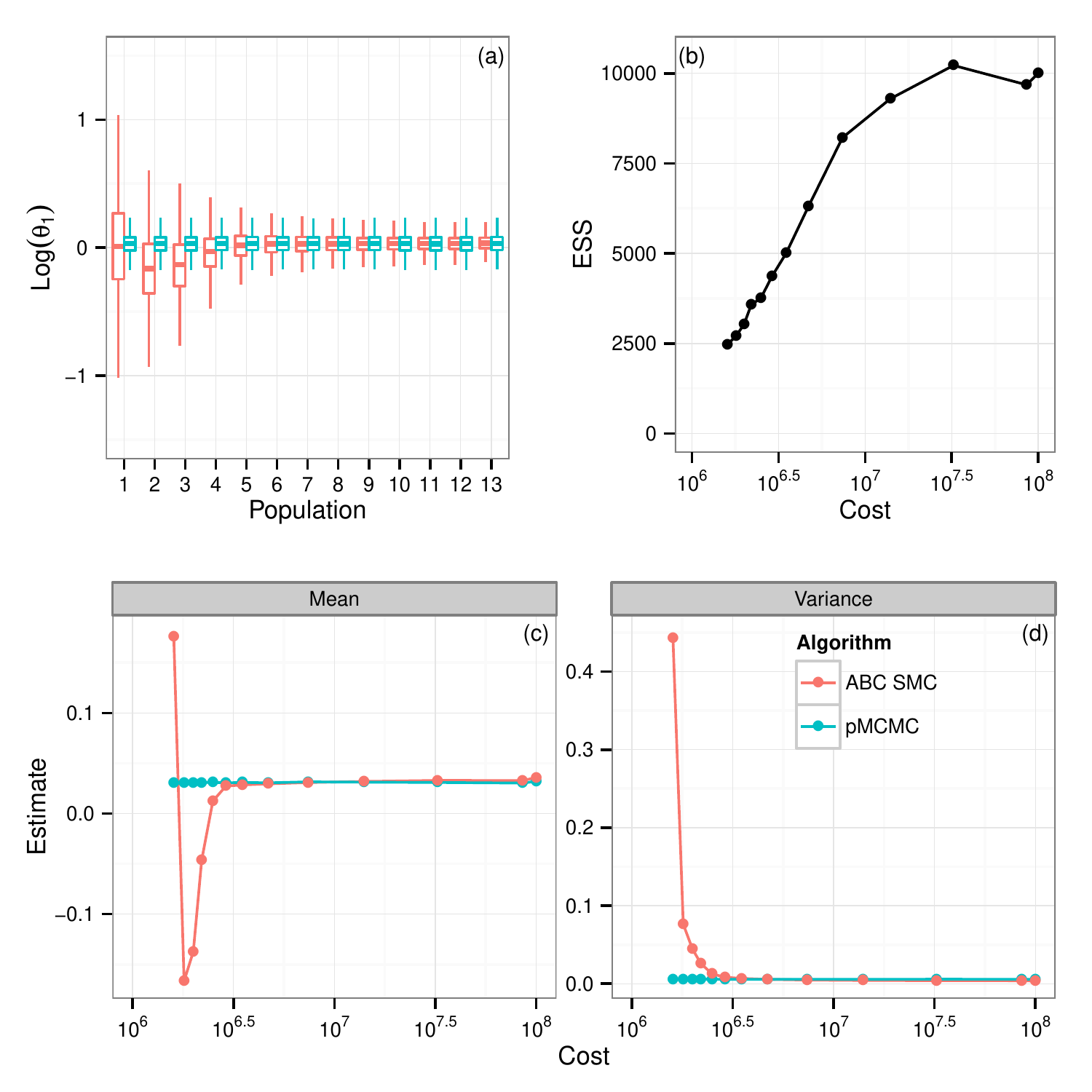}
  \caption{Further investigation about the properties of each posterior sample using a given run from each algorithm $\mathcal{D}^{1}$. Plots all concern the first rate parameter. (a) are box plots of the distribution factored into computational expense according to the series of distributions in the ABC SMC. (b) shows effective sample size of the pMCMC posterior given these cost brackets. (c) and (d) are posterior estimates of mean and variance given one run of each algorithm.}
  \label{fig:lvsmallproperties}
\end{figure}

The differences between pMCMC and ABC SMC inference are less pronounced for $\mathcal{D}^{1}$ and $\mathcal{D}^{1}_{u}$ particularly for the reaction rate parameters, see figure~\ref{fig:lvsmallposterior}. The analysis shown in figure~\ref{fig:lvsmallproperties} of $\mathcal{D}^{1}$ imply that the posterior inferences made after a short time, less than 10\% of the total allocated budget, are largely indistinguishable between the two approaches, estimates of the mean and variance are equivalent and the shape of the distribution consistent between the two approaches. If we take into consideration the cost of initialising pMCMC this would suggest that in this case ABC SMC proves a more favorable choice. However this effect is less pronounced when some information is removed. Posterior learning of the noise parameter appears to be poor, consistent with results in the text, but this is now also true for pMCMC. However the pMCMC runs subject to the budget constraints are in agreement with the long run, black, being used as the truth for comparison. This ``truth'' being the results of a long well thinned run using a large number of particles. 

\subsection{Analysing Lotka--Volterra $\mathcal{D}^{3}_{\ast}$}
\label{ssec:supp2}

\begin{figure}[!h]
  \centering
  \includegraphics[width = \textwidth]{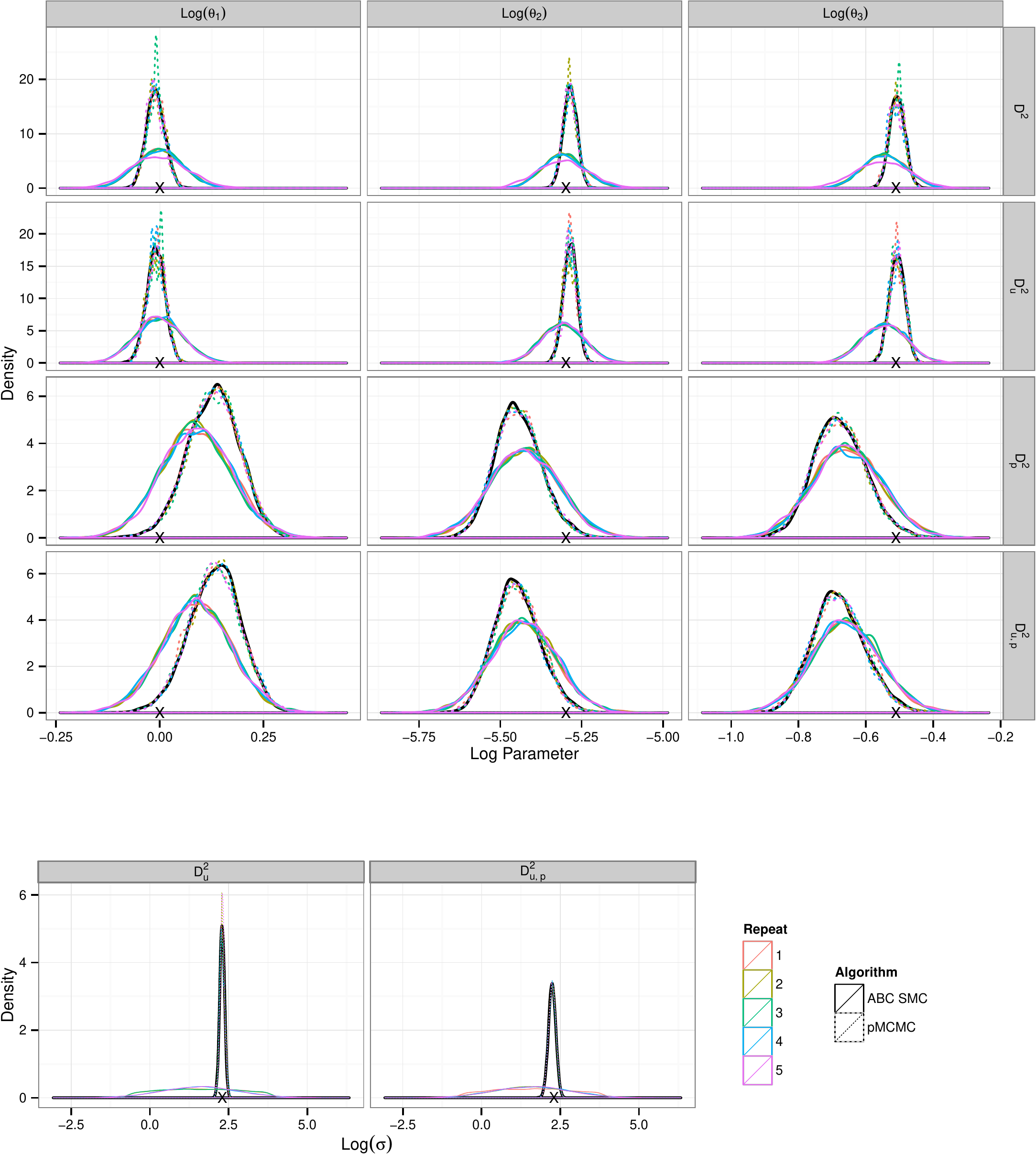}
  \caption{Posterior densities for the $\mathcal{D}^{3}_{\ast}$ collection of data sets using each algorithm for each of the data observation regimes. ABC SMC puts more emphasis on the tails of the distribution. True posterior distributions are shown in black.}
  \label{fig:lvlargeposteriors}
\end{figure}

\begin{figure}[!h]
  \centering
  \includegraphics[width = \textwidth]{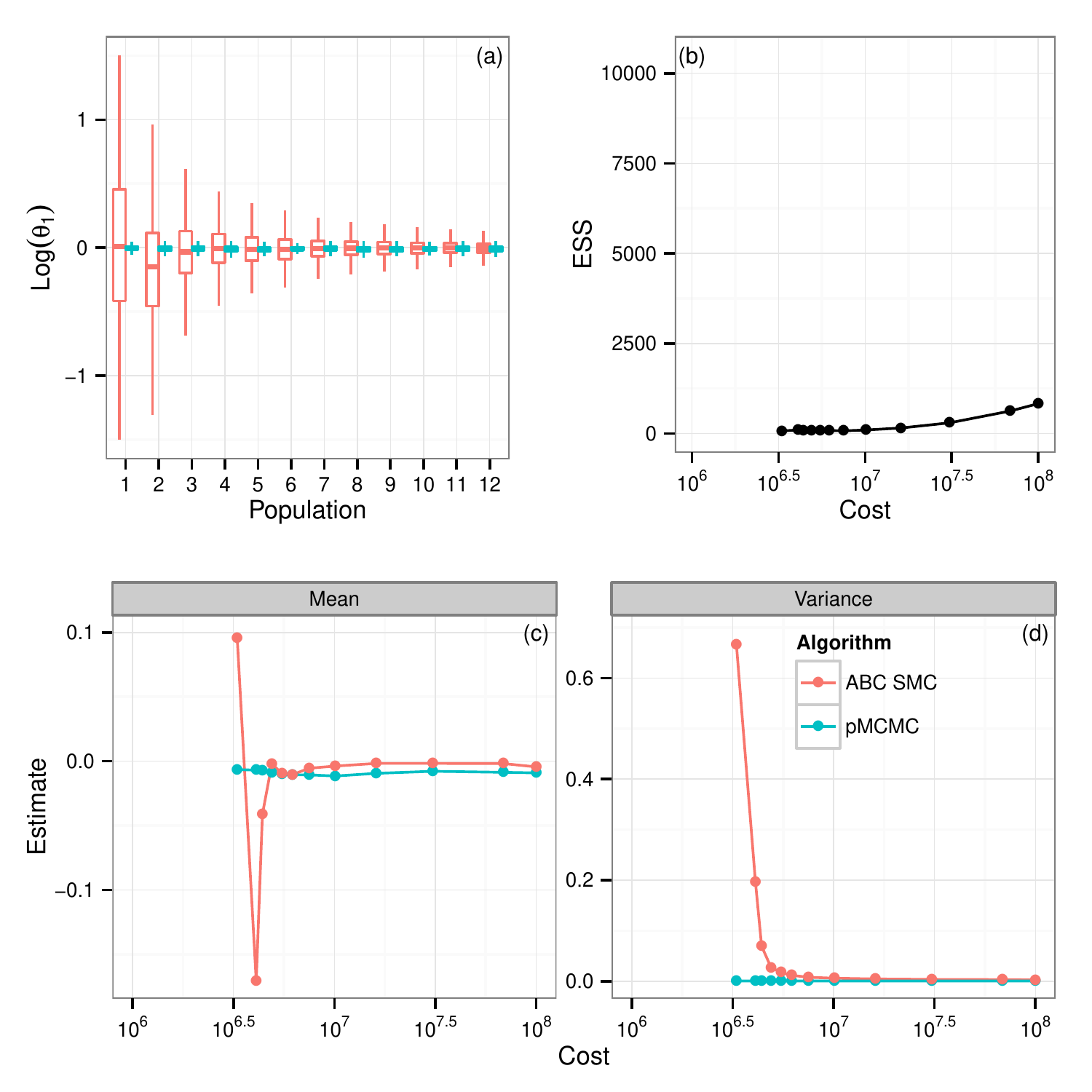}
  \caption{Posterior summary information for $\mathcal{D}^{3}$ $\theta_{1}$ segmented into computational groups. (a) shows that the ABC distributions are tending toward the pMCMC results. (b) shows that the effective sample size of pMCMC under this computational restriction is small. (c) and (d) show that posterior estimates of mean and variance are similar for both algorithms.}
  \label{fig:lvlargeproperties}
\end{figure}

Inference for the larger data sets shows a larger discrepancy between the two algorithms, see figure~\ref{fig:lvlargeposteriors} for density plots. ABC SMC consistently over--estimates the contribution in the tails of the distribution. With more abundant data pMCMC infers the noise parameter, $\sigma$, well whilst ABC SMC continues to do poorly in this respect, consistent with other results in this article. In all data observation regimes pMCMC results in a more peaked posterior distribution that closely matches our ``truth'' benchmark. Figure~\ref{fig:lvlargeproperties}(a) shows that ABC SMC is tending in distribution to that we found by using pMCMC. Figure~\ref{fig:lvlargeproperties} shows that obtaining a diverse posterior sample using pMCMC is more computationally taxing for the larger dimensional data. Despite the difference between the posterior densities, estimates of the means and variances appear consistent after approximately 10\% of the allowed budget.

\end{document}